\documentclass[prd,aps,showpacs,preprintnumbers,amssymb]{revtex4}

\usepackage{color}
\usepackage{epsf}

\begin{document}
\title{%
\hfill{\normalsize\vbox{%
\hbox{}
 }}\\
{An approach to QCD perturbative and non-perturbative beta functions }}

\author{Renata Jora
$^{\it \bf a}$~\footnote[2]{Email:
 rjora@theory.nipne.ro}}

\affiliation{$^{\bf \it a}$ National Institute of Physics and Nuclear Engineering PO Box MG-6, Bucharest-Magurele, Romania}

\date{\today}

\begin{abstract}
Using the global properties of the QCD partition function we determine  an all order perturbative beta function in the background gauge field method to find out that it has a simple expressions whose properties and consequences align with other recent results in the QCD literature. We further compute the non-perturbative beta functions for the coupling constant and the theta angle in the background of an instanton field with winding number $n$.  We solve for the theta angle  in the non-perturbative region to determine that it is approximately zero. By extrapolating to the full QCD beta functions our result may constitute a solution to the strong CP problem.
\end{abstract}
\maketitle

\section{Introduction}
The behavior of quantum chromodynamics is of utmost interest both at high energies where the coupling constant is small and perturbation theory makes sense and at low energies where the coupling constant is large and quarks form bound states, the hadrons.  The perturbative beta function in QCD has been computed up to the fifth order in the minimal regularization scheme \cite{Larin1},\cite{Larin2}, \cite{Baikov}. In the background gauge field method beta function for an arbitrary Yang Mills theory with fermions has been calculated up to five loops in \cite{Abbott1}, \cite{Abbott2}, \cite{Vermaseren}. It is well known that the first two orders coefficients of the beta function are renormalization scheme independent and as such there exist a renormalization scheme where the beta function stops at the first two orders \cite{Hooft1}, \cite{Hooft2},  \cite{Jora}. However it is argued that such scheme may pose difficulties with computing other quantities in the renormalized perturbation theory. It is quite generic that if only the first two coefficients of the beta function are renormalization scheme independent then one might obtain for higher orders at least in principle any well behaved function. Knowing the exact beta function  is as important as the scheme that leads to it because one needs to approach all phenomena in the same context. In \cite{Sannino} Pica and Sannino proposed an all order beta function for the the QCD coupling constant inspired by the famous NSVZ $N=1$ supersymmetric beta function \cite{Novikov}. The Yang Mills version of this has a simple formula as:
\begin{eqnarray}
\beta_{YM}=-\frac{11}{3}\frac{\alpha^2}{2\pi}\frac{C_2(G)}{1-\frac{\alpha}{2\pi}\frac{17}{11}C_2(G)},
\label{sanbeta}
\end{eqnarray}
where $\alpha=\frac{g^2}{16\pi^2}$ and $C_2(G)=N$ where $N$ is the number of colors.
In the presence of fermions the Pica -Sannino beta function is modified to:
\begin{eqnarray}
\beta(\alpha)=-\frac{\alpha^2}{2\pi}\frac{a+\sum_{r=1}^pa_rN_r\gamma_r}{1-\frac{\alpha}{2\pi}a_g}
\label{snabeta25890}
\end{eqnarray}
Here $\alpha=\frac{g^2}{4\pi}$ and,
\begin{eqnarray}
&&\beta(\alpha)=\frac{d \alpha}{d \ln\mu}
\nonumber\\
&&\gamma_r=-\frac{d\ln m_r}{d\ln\mu},
\label{res756677}
\end{eqnarray}
where $\beta(\alpha)$ is the beta function of the coupling constant and $\gamma_r$ is the anomalous dimension of the fermion mass operator. The coefficients $a$, $a_r$ and $a_g$ are further determined by matching  the first two order coefficients renormalization scheme independent of the beta function.

Regular Yang Mills and QCD are by far more complicated models than their supersymmetric counterparts that have been discussed and elucidated in a series of groundbreaking works in \cite{Seiberg1}-\cite{Seiberg3}. In \cite{Seiberg3} Seiberg showed that at least for the $N=2$ supersymmetric gauge theories the perturbative beta function stops at one loop and also introduced the first order non-perturbative effects for both the coupling constant and the theta angle.   Of topical interest is not only the beta function for the coupling constant but also the behavior of the theta angle with the scale. Although early attempts have been made to extract at least the contribution of the theta angle to the coupling constant beta function \cite{Callan}, or the form of both beta functions with the scale \cite{Morozov} not definite answer has emerged in this direction. But the exact behavior of the the theta angle with the scale and the strong coupling constant is crucial as it might lead to insights with regard to the strong CP problem.

In this work we shall consider QCD  in the background gauge field theory and based on the properties of the partition function  we shall derive the exact form of the complete beta function in the perturbative approach to determine that its expression has some similarities but also differences with that suggested by Pica and Sannino in \cite{Sannino}. Moreover we shall also determine non-perturbative effect in the background  of an arbitrary instanton field. In this context we estimate also the beta function for the theta angle. Our results have some common features with the Seiberg non-perturbative supersymmetric beta function and to other approaches but also significant differences. Based on the two complete beta function for the coupling constant and the theta angle we extract the behavior of the theta angle with the coupling constant at low energies where the coupling constant is large. We find that in this regime the theta angles runs to the effective value a zero. This results is in striking agreement with the experimental results like that for electric dipole momentum for the neutrons \cite{Afach} that suggest that the theta angle is close to zero so it might constitute a solution to the strong CP problem.

In Section II we present our set-up.  Sections III and IV contain our method and the main derivations. In section V we  determine the perturbative beta function whereas in section VI we discuss the non-perturbative beta functions for the coupling constant and the theta angle. Section VII is dedicated to Conclusions.

\section{The set-up}
 We start from the QCD Lagrangian in the background gauge field method \cite{Peskin}:
 \begin{eqnarray}
&& {\cal L}=-\frac{1}{4g^2}[F^a_{\mu\nu}+D_{\mu}A^a_{\nu}-D_{\nu}A^a_{\mu}+f^{abc}A^b_{\mu}A^c_{\nu}]^2-
 \nonumber\\
 &&-\frac{1}{2g^2}(D^{\mu}A^a_{\mu})^2-\bar{c}^a(D^2)^{ac}c^c-\bar{c}^aD^{\mu}f^{abc}A^b_{\mu}c^c+\sum_f\bar{\Psi}_fi\gamma^{\mu}(\partial_{\mu}-igt^aA^a_{\mu})\Psi_f,
 \label{lagr6574}
 \end{eqnarray}
 where,
 \begin{eqnarray}
 A^a_{\mu}\rightarrow B^a_{\mu}+A^a_{\mu}
 \label{dec54}
 \end{eqnarray}
 and $B^a_{\mu}$ is the background gauge field and $A^a_{\mu}$ in the decomposition is the quantum gauge field fluctuation. Also,
 \begin{eqnarray}
 (D_{\mu})^{ac}=\partial_{\mu}\delta^{ac}+B^b_{\mu}f^{abc},
 \label{covder}
 \end{eqnarray}
 where $D_{\mu}$ is the covariant derivative in the background gauge field method that it is applied to the quantum gauge field and to the ghosts.

 Next we shall consider the partition function associated with QCD in the background gauge field method:
 \begin{eqnarray}
 Z=\int d A^a_{\mu} d \bar{c}^b d c^b d\bar{\Psi}_f d \Psi_f\exp[i\int d^4 x {\cal L}(B,A,c)].
 \label{partfunc657}
 \end{eqnarray}
 We allow for the possibility of a theta angle term and extend this partition function to:
 \begin{eqnarray}
 Z_{\theta}=\int d A^a_{\mu} d \bar{c}^b d c^b d\bar{\Psi}_f d \Psi_f\exp[i\int d^4 x {\cal L}(B,A,c)+i \theta\int d^4x{\cal L}_1(B,A,c)]
 \label{theta546738}
 \end{eqnarray}
 where,
 \begin{eqnarray}
 {\cal L}_1=\frac{1}{32\pi^2}\epsilon^{\mu\nu\rho\sigma}[F^a_{\mu\nu}+D_{\mu}A^a_{\nu}-D_{\nu}A^a_{\mu}+f^{abc}A^b_{\mu}A^c_{\nu}][F^a_{\rho\sigma}+D_{\rho}A^a_{\sigma}-D_{\sigma}A^a_{\rho}+f^{abc}A^b_{\rho}A^c_{\sigma}].
 \label{theta42634}
 \end{eqnarray}
 The calculation in the background gauge field are made in the saddle approximation where the linear term is set to zero. Since this is equivalent to asking that the background gauge field satisfies the equation of motion we shall consider this as a constraint that applies to all the calculations in the present work.

 One can integrate the partition function to obtain the effective potential whose form is known in the first orders and envisioned in all orders:
 \begin{eqnarray}
&& Z_{\theta,B}=\exp[-\Gamma_{eff}]=
 \nonumber\\
 &&\exp[-\int d^4 x \frac{1}{4g^2(\mu)}F^a_{\mu\nu}F^{a\mu\nu}+i\theta(\mu)\frac{1}{32\pi^2}\int d^4 x \tilde{F}^{a\mu\nu}F^a_{\mu\nu}+{\rm higher\,\,order\,\,gauge\,\,invariants}],
 \label{finalres746657}
 \end{eqnarray}
where the equality stands up to some constant derivative factors.

 Next we shall make a change of variables in the integral for the partition function given in Eq. (\ref{theta546738}) back to the original gauge field $A^a_{\mu}\rightarrow A^a_{\mu}-B^a_{\mu}$. Then we  retrieve the original Lagrangian that does not depend on the gauge field plus an extra term gauge dependent according to:
 \begin{eqnarray}
&& {\cal L}_{B,A,c,\Psi}\rightarrow {\cal L}_{B=0}+i\theta({\cal L}_{B=0})_1+{\cal L}_B
 \nonumber\\
&&{\cal L}_B= -\bar{c}^af^{abc}B^b_{\mu}\partial_{\mu}c^c-\bar{c}^af^{amn}f^{nbc}B^m_{\mu}A^b_{\mu}c^c.
 \label{res64536478}
 \end{eqnarray}
 Then Eqs. (\ref{theta546738}), (\ref{finalres746657}) and (\ref{res64536478}) lead to the following master formula which will be used in everything that follows:
 \begin{eqnarray}
&&\exp[-\int d^4 x \frac{1}{4g^2(\mu)}F^a_{\mu\nu}F^{\mu\nu}+i\theta(\mu)\frac{1}{32\pi^2}\int d^4 x \tilde{F}^{a\mu\nu}F^a_{\mu\nu}+{\rm higher\,\,order\,\,gauge\,\,invariants}]=
\nonumber\\
&&\int d A^a_{\mu} d\bar{c}^b dcd\bar{\Psi}_f d\Psi_f\exp[\int d^4 x[{\cal L}_{B=0}+i\theta({\cal L}_{B=0})_1+{\cal L}_B]].
\label{finalremaster4566}
\end{eqnarray}
In the end we shall make the notation:
\begin{eqnarray}
\int d^4 x[{\cal L}_{B=0}+i\theta({\cal L}_{B=0})_1]=S(A,c,\Psi,\theta).
\label{notrytuur6575}
\end{eqnarray}

\section{Method I}
In this section we will set the theta term to zero and consider a  background gauge field  which satisfies the equation of motion.
We apply the operators $\frac{\delta^2}{\delta B^m_{\rho}(z)\delta B^m_{\rho}(z)}$ and  $\frac{\delta^4}{\delta B^m_{\rho}(z)\delta B^m_{\rho}(z)\delta B^{m_1}_{\sigma}(z)\delta B^{m_1}_{\sigma}(z)}$to the left and right hand side of the Eq. (\ref{finalremaster4566}).
 First we shall consider the right hand side:
 \begin{eqnarray}
&&W_1= \frac{\delta^2}{\delta B^m_{\rho}(z)\delta B^{m\rho}(z)}\int d A^a_{\mu} d\bar{c}^b dc^bd\bar{\Psi}_f d\Psi_f\exp[\int d^4 x[{\cal L}_{B=0,A,c}+i\theta({\cal L}_{B=0,A,c})_1+{\cal L}_B]]=
 \nonumber\\
 &&= \int d A^a_{\mu} d\bar{c}^b d\bar{\Psi}_f d\Psi_f dc^b[\bar{c}^a(z)f^{amc}\partial_{\rho}c^c(z)+\bar{c}^af^{amn}f^{nbc}A^b_{\rho}c^c][\bar{c}^{a_1}(z)f^{a_1mc_1}\partial^{\rho}c^{c_1}(z)+\bar{c}^{a_1}f^{a_1mn_1}f^{n_1b_1c_1}A^{b_1\rho}c^{c_1}]
 \times\nonumber\\
 &&\exp[\int d^4 x[{\cal L}_{B=0,A,c}+i\theta({\cal L}_{B=0,A,c})_1+{\cal L}_B]].
 \label{fomr65746575}
 \end{eqnarray}
Since in the end we shall consider the value of this differential operator for $B=0$ we observe (for the simplicity of the relations we shall omit all the indices knowing that the summation over them will lead to dimensionless coefficients whose specific values are irrelevant here):
\begin{eqnarray}
W_1={\rm const} \int dA^a_{\mu}d\bar{c}^b dc^b \bar{c}(z)d\bar{\Psi}_f d\Psi_f\bar{c}(z)\frac{\delta}{\delta(\partial_{\mu}\bar{c}(z))}\frac{\delta}{\delta(\partial^{\mu}\bar{c}(z))}\exp[\int d^4 x[{\cal L}_{B=0,A,c}+i\theta({\cal L}_{B=0,A,c})_1+{\cal L}_B]]|_{B=0}.
\label{res6453424}
\end{eqnarray}
where color indices of the ghost fields in the product are different.
This is evident because the ghost term that contains B is of the form $-\bar{c}B^{\mu}D_{\mu}c$  whereas that that appears in the free action is just $\partial_{\mu}(\bar{c})D_{\mu}c$.

Next we consider:
\begin{eqnarray}
&&W_2=\frac{\delta^4}{\delta B^m_{\rho}(z)\delta B^{m\rho}(z)\delta B^{m_1}_{\sigma}(z)\delta B^{m_1\sigma}(z)}\int d A^a_{\mu} d\bar{c}^b dc^b d\bar{\Psi}_f d\Psi_f\exp[\int d^4 x[{\cal L}_{B=0,A,c}+i\theta({\cal L}_{B=0,A,c})_1+{\cal L}_B]]|_{B=0}=
\nonumber\\
&&= \int d A^a_{\mu} d\bar{c}^b dc^b d\bar{\Psi}_f d\Psi_f \times\nonumber\\
&&\Bigg[\bar{c}^a(z)f^{amc}\partial_{\rho}c^c(z)+\bar{c}^af^{amn}f^{nbc}A^b_{\rho}c^c][\bar{c}^{a_1}(z)f^{a_1mc_1}\partial^{\rho}c^{c_1}(z)+\bar{c}^{a_1}f^{a_1mn_1}f^{n_1b_1c_1}A^{b_1\rho}c^{c_1}]\times
\nonumber\\
&&[\bar{c}^{a_2}(z)f^{a_2m_1c_2}\partial_{\sigma}c^{c_2}(z)+\bar{c}^{a_2}f^{a_2m_1n_2}f^{n_2b_2c_2}A^{b_2}_{\sigma}c^{c_2}]
[\bar{c}^{a_3}(z)f^{a_3m_1c_3}\partial^{\sigma}c^{c_3}(z)+\bar{c}^{a_3}f^{a_3m_1n_3}f^{n_3b_3c_3}A^{b_3\sigma}c^{c_3}]
 \nonumber\\
 &&\exp[\int d^4 x[{\cal L}_{B=0,A,c}+i\theta({\cal L}_{B=0,A,c})_1+{\cal L}_B]]_{B=0}.
 \label{intfirst657444}
 \end{eqnarray}
Based on the same arguments as before one can write:
\begin{eqnarray}
&&W_2={\rm const'}\int d A^a_{\mu} d\bar{c}^b dc^b d\bar{\Psi}_f d\Psi_f\bar{c}(z)\bar{c}(z)\bar{c}(z)\bar{c}(z)\times
\nonumber\\
&&\frac{\delta}{\delta(\partial_{\mu}\bar{c}(z))}\frac{\delta}{\delta(\partial^{\mu}\bar{c}(z))}\frac{\delta}{\delta(\partial_{\rho}\bar{c}(z))}\frac{\delta}{\delta(\partial^{\rho}\bar{c}(z))}
\exp[\int d^4 x[{\cal L}_{B=0,A,c}+i\theta({\cal L}_{B=0,A,c})_1+{\cal L}_B]]=
\nonumber\\
&&{\rm const''}\int d A^a_{\mu} d\bar{c}^b dc^b d\bar{\Psi}_f d\Psi_f\frac{\delta}{\delta(\partial_{\mu}\bar{c}(z))}\frac{\delta}{\delta(\partial^{\mu}\bar{c}(z))}\times
\nonumber\\
&&\Bigg[\bar{c}(z)\bar{c}(z)\bar{c}(z)\bar{c}(z)\frac{\delta}{\delta(\partial_{\rho}\bar{c}(z))}
\frac{\delta}{\delta(\partial^{\rho}\bar{c}(z))}\exp[\int d^4 x[{\cal L}_{B=0,A,c}+i\theta({\cal L}_{B=0,A,c})_1+{\cal L}_B]]\Bigg]-
\nonumber\\
&&{\rm const''}\int d A^a_{\mu} d\bar{c}^b dc^b d\bar{\Psi}_f d\Psi_f\Bigg[\frac{\delta}{\delta(\partial_{\mu}\bar{c}(z))}\frac{\delta}{\delta(\partial^{\mu}\bar{c}(z))}
\bar{c}(z)\bar{c}(z)\Bigg]\times
\nonumber\\
&&\bar{c}(z)\bar{c}(z)\frac{\delta}{\delta(\partial_{\rho}\bar{c}(z))}\frac{\delta}{\delta(\partial_{\rho}\bar{c}(z))}
\exp[\int d^4 x[{\cal L}_{B=0,A,c}+i\theta({\cal L}_{B=0,A,c})_1+{\cal L}_B]]=
\nonumber\\
&&{\rm const''}\int d A^a_{\mu} d\bar{c}^b dc^b d\bar{\Psi}_f d\Psi_f\Bigg[\frac{\delta}{\delta(\partial_{\mu}\bar{c}(z))}\frac{\delta}{\delta(\partial^{\mu}\bar{c}(z))}
\bar{c}(z)\bar{c}(z)\Bigg]\times
\nonumber\\
&&\bar{c}(z)\bar{c}(z)\frac{\delta}{\delta(\partial_{\rho}\bar{c}(z))}\frac{\delta}{\delta(\partial^{\rho}\bar{c}(z))}
\exp[\int d^4 x[{\cal L}_{B=0,A,c}+i\theta({\cal L}_{B=0,A,c})_1+{\cal L}_B]],
\label{calc6574}
\end{eqnarray}
where we used the principles of integration by parts that work as well for noncommutative variables. Next we need to determine the quantity:
\begin{eqnarray}
&&\Bigg[\frac{\delta}{\delta(\partial_{\mu}\bar{c}(z))}\frac{\delta}{\delta(\partial^{\mu}\bar{c}(z))}\bar{c}(z)\bar{c}(z)\Bigg]=
\nonumber\\
&&\frac{\delta}{\delta(\partial_{\mu}\bar{c}(z))}\frac{\delta}{\delta(\partial^{\mu}\bar{c}(z))}[\int dy\bar{c}(y)\bar{c}(y)\delta(y-z)]=
\nonumber\\
&&-\frac{\delta}{\delta(\partial_{\mu}\bar{c}(z))}\frac{\delta}{\delta(\partial^{\mu}\bar{c}(z))}[\int dy \bar{c}(y)\bar{c}(y)(\partial_{\rho})^2\int \frac{d^4p}{(2\pi)^4}\frac{1}{p^2}\exp[ip(y-z)]]\propto
\nonumber\\
&&-\frac{\delta}{\delta(\partial_{\mu}\bar{c}(z))}\frac{\delta}{\delta(\partial^{\mu}\bar{c}(z))}[\int dy [(\partial_{\rho})^2\bar{c}(y)\bar{c}(y)+\bar{c}(y)(\partial_{\rho})^2\bar{c}(y)+
2\partial_{\rho}\bar{c}(y)\partial^{\rho}\bar{c}(y)]\times
\nonumber\\
&&\int \frac{d^4p}{(2\pi)^4}\frac{1}{p^2}\exp[ip(y-z)]=
-4\int\frac{d^4k}{(2\pi)^4}\int \frac{d^4p}{(2\pi)^4}\frac{1}{p^2}.
\label{res54112}
\end{eqnarray}
Here we used:
\begin{eqnarray}
&&\frac{\delta(\partial_{\rho}\bar{c}(y))}{\delta (\partial_{\mu}\bar{c})}=\delta_{\mu\rho}\delta(z-y)
\nonumber\\
&&\frac{\delta(\partial_{\rho}\partial_{\rho}\bar{c}(y))}{\delta (\partial_{\mu}\bar{c})}=\partial_{\rho}\delta(z-y)
\nonumber\\
&&\int dx f(x) \delta'(a-x)=f(a)'
\label{prop776}
\end{eqnarray}
where the last equation in Eq. (\ref{prop776})  is a generic property of the delta functions.
By combining Eq. (\ref{calc6574}) and Eq. (\ref{res54112}) we obtain:
\begin{eqnarray}
W_2=xW_1\times\int\frac{d^4k}{(2\pi)^4}\int \frac{d^4p}{(2\pi)^4}\frac{1}{p^2}
\label{res74536}
\end{eqnarray}
where $x$ is a dimensionless constant that depends on the group factors and constants.

Next we need to compute the same derivatives for the left hand side of the Eq. (\ref{finalremaster4566}). Since in the end we will set the field $B$ to zero the results of these derivatives will be:
\begin{eqnarray}
&& \frac{\delta^2}{\delta B^m_{\rho}(z)\delta B^{m\rho}(z)}\exp[-\Gamma(B)]|_{B=0}=a_0 \frac{\delta^2 \Gamma(B)}{\delta B^m_{\rho}(z)\delta B^{m\rho}(z)}|_{B=0}
 \nonumber\\
 &&\frac{\delta^4}{\delta B^m_{\rho}(z)\delta B^{m\rho}(z)\delta B^{m_1}_{\sigma}(z)\delta B^{m_1\sigma}(z)}\exp[-\Gamma(B)]|_{B=0}=
 \nonumber\\
 &&a_1\frac{\delta^4 \Gamma[B]}{\delta B^m_{\rho}(z)\delta B^{m\rho}(z)\delta B^{m_1}_{\sigma}(z)\delta B^{m_1\sigma}(z)}+
 \nonumber\\
&& a_2\frac{\delta^2 \Gamma[B]}{\delta B^m_{\rho}\delta B^{m\rho}}\frac{\delta^2 \Gamma[B]}{\delta B^{m_1}_{\sigma} \delta B^{m_1\sigma}}+
 \nonumber\\
&& a_3\frac{\delta^2 \Gamma[B]}{\delta B^m_{\rho}\delta B^{m_1\sigma}}\frac{\delta^2 \Gamma[B]}{\delta B^{m_1}_{\sigma} \delta B^{m\rho}}+{\rm similar\,\,terms\,\,to\,\,the\,\,previous\,\,ones}.
 \label{resy67588}
 \end{eqnarray}
 Here the coefficients $a_0$, $a_1$, $a_2$ and $a_3$ are dimensionless and since we do not aim to compute these relation exactly the dependence on the space time and internal indices is irrelevant. However the dependence on the variable space time is crucial.
 Then,
 \begin{eqnarray}
&& \frac{\delta^2 \Gamma(B)}{\delta B^m_{\rho}(z)\delta B^{m\rho}(z)}|_{B=0}\propto
\nonumber\\
&&  \frac{\delta^2}{\delta B^m_{\rho}(z)\delta B^{m\rho}(z)}[\int d^4 x\int d^4 y \frac{d^4k}{(2\pi)^4}\frac{1}{2g^2}B^a_{\mu}(x)\exp[-ikx][-k^2g^{\mu\nu}+k^{\mu}k^{\nu}]B^a_{\nu}(y)\exp[iky]] =
\nonumber\\
&&\frac{1}{2g^2}\int d^4 xd^4 y \frac{d^4k}{(2\pi)^4}\delta(x-z)\delta(y-z)\delta_{\mu\nu}\exp[-ikx+iky][-k^2g^{\mu\nu}+k^{\mu}k^{\nu}]=
\nonumber\\
&&\frac{b_0}{g^2}\int \frac{d^4k}{(2\pi)^4}k^2,
\label{res64553}
\end{eqnarray}
where the summation over space time indices is understood. Also $b_0$ is a dimensionless numerical coefficient irrelevant for our purposes.
Moreover,
\begin{eqnarray}
&&\frac{\delta^4 \Gamma[B]}{\delta B^m_{\rho}(z)\delta B^{m\rho}(z)\delta B^{m_1}_{\sigma}(z)\delta B^{m_1\sigma}(z)}=
\nonumber\\
&&\frac{1}{g^2}\frac{\delta^4 }{\delta B^m_{\rho}(z)\delta B^{m\rho}(z)\delta B^{m_1}_{\sigma}(z)\delta B^{m_1\sigma}(z)}\int d^4 x f^{abc} f^{ade} B^b_{\mu}(x)b^c_{\nu}(x)B^d_{\mu}(x)B^e_{\mu}(x)=
\nonumber\\
&&\frac{1}{g^2}b_1\int d^4x \delta(x-z)\delta(x-z)\delta(x-z)\delta(x-z)=b_1\frac{1}{g^2}[\int \frac{d^4p}{(2\pi)^4}]^3.
\label{second65774}
\end{eqnarray}
Here the coefficient $b_1$ is dimensionless but its value is irrelevant for our purposes. The contributions of the terms with coefficients $a_2$ and $a_3$ in the quadrilinear derivative in Eq. (\ref{resy67588}) can be determined easily from the square of the results in Eq. (\ref{res64553}) only with different dimensionless coefficients. Then one may write:
\begin{eqnarray}
&&\frac{\delta^4}{\delta B^m_{\rho}(z)\delta B^{m\rho}(z)\delta B^{m_1}_{\sigma}(z)\delta B^{m_1\sigma}(z)}\exp[-\Gamma(B)]|_{B=0}=
\nonumber\\
&&b_1\frac{1}{g^2}[\int \frac{d^4p}{(2\pi)^4}]^3+b_2\frac{1}{g^4}[\int \frac{d^4k}{(2\pi)^4}k^2]^2,
\label{finresq5463}
\end{eqnarray}
where again $b_2$ is a dimensionless coefficient.

Next using Eqs (\ref{res74536}), (\ref{res64553}), (\ref{second65774}) and (\ref{finresq5463}) we determine for the quadrilinear derivative  of the Eq. (\ref{finalremaster4566}):
\begin{eqnarray}
a\frac{1}{g^2}[\int \frac{d^4p}{(2\pi)^4}]^3+b\frac{1}{g^4}[\int \frac{d^4k}{(2\pi)^4}k^2]^2=-c\frac{1}{g^2}[\int \frac{d^4k}{(2\pi)^4}k^2]\int\frac{d^4p}{(2\pi)^4}\frac{1}{p^2}\int \frac{d^4r}{(2\pi)^4}],
\label{fin564555}
\end{eqnarray}
where $a$, $b$ and $c$ are dimensionless coefficients that depend on the group constants.

We shall consider the expression in Eq. (\ref{fin564555}) in the dimensional regularization scheme (see  Appendix A for the explicit calculations).  Then according to our calculations and notations Eq. (\ref{res74536}) will become:
\begin{eqnarray}
Z_0W_{2r}=xZ_0W_{1r}\times\int\frac{d^dk}{(2\pi)^d}\int \frac{d^dp}{(2\pi)^d}\frac{1}{p^2},
\label{re64775885}
\end{eqnarray}
where $W_{2r}$ and $W_{1r}$ denote the derivatives associated to the right hand side and the following relations hold:
\begin{eqnarray}
&&W_{2r}=Z_0^{-1}W_{2l}
\nonumber\\
&&W_{1r}=Z_0^{-1}W_{1l},
\label{newrel657884}
\end{eqnarray}
Here $W_{2l}$ and $W_{1l}$ represent the derivatives associated to the left hand side. Consequently  Eq. (\ref{re64775885}) will become:
\begin{eqnarray}
W_{2l}=xW_{1l}\times\int\frac{d^dk}{(2\pi)^d}\int \frac{d^dp}{(2\pi)^d}\frac{1}{p^2}.
\label{res5524439}
\end{eqnarray}
Considering all these aspects Eq. (\ref{fin564555}) will  be expressed as:
\begin{eqnarray}
\Bigg[a\frac{1}{g^2\mu^{2\epsilon}}[\frac{x_1}{\epsilon}+x_0+...]+b\frac{1}{g^4\mu^{4\epsilon}}+c\frac{1}{g^2\mu^{2\epsilon}}[\frac{z_1}{\epsilon}+z_0+...]\Bigg]Z_0=0,
\label{fires534442}
\end{eqnarray}
where $Z_0$ is the partition function in the absence of the background gauge field. We can rewrite Eq. (\ref{fires534442}) as:
\begin{eqnarray}
\Bigg[\frac{1}{g^2}\mu^{2\epsilon}(\frac{r_1}{\epsilon}+r_0+...)+2b\frac{1}{g^4}\Bigg]Z_0=0.
\label{res664553}
\end{eqnarray}
Here:
\begin{eqnarray}
&&r_1=ax_1+cz_1
\nonumber\\
&&r_0=ax_0+cz_0.
\label{rel856677}
\end{eqnarray}

\section{Method II}

Here we shall determine the properties and the behaving with the scale of the $Z_0$, the partition function in the absence of the background gauge field.

The first step is  to establish that the result of integrating in the partition function will lead to:
\begin{eqnarray}
&&\prod_{f,i,a}\int d A^a_{\mu}d\bar{\Psi}_{fi}d\Psi_{fi}d\bar{c}_adc_a\exp[\int d^4 x{\cal L}]\approx
\nonumber\\
&&\prod_k\det[\frac{1}{k^2}]^{2(N^2-1)}\det[\frac{1}{k^2}]^{-(N^2-1)}\det[\frac{1}{k^2}]^{-2N_fN}\exp[iS_0]
\label{resofintegrt455353}
\end{eqnarray}
Here $f$ is the fermion flavor, $i$ is the color component and $a$ is the color index in the adjoint representation. Moreover we extracted only the nontrivial contributions that depend on the momenta and also for the fermion case we used $\prod_k\det \frac{1}{\gamma^{\mu}k_{\mu}}\approx\prod_{k>0}\frac{1}{k^2}=[\prod_k\frac{1}{k^2}]^{1/2}$.
Next we use (see \cite{Peskin}):
\begin{eqnarray}
\det[\frac{1}{p^2}]^a\approx\exp[iq\int d^d x\int\frac{d^dp}{(2\pi)^d}\ln[-p^2+m^2]],
\label{res6352442}
\end{eqnarray}
where $m$ is a small universal infrared regulator. In order to compute the expression in Eq. (\ref{res6352442}) we use:
\begin{eqnarray}
&&\int \frac{d^dp}{(2\pi)^d}\ln[-p^2+m^2]=-i\Gamma[-d/2]\frac{1}{(4\pi)^{d/})}(m^2)^{d/2}
\nonumber\\
&&\int d^d x=-i\int d^d x_E \frac{x_E^2}{x_E^2+\frac{\epsilon^{4/d}}{m^2}}=-i\pi^{d/2}\epsilon^2\frac{d}{2}(m^2)^{-d/2}\Gamma[-d/2].
\label{somecalc774665}
\end{eqnarray}
Then the result of computing the product of determinants in Eq. (\ref{resofintegrt455353}) is:
\begin{eqnarray}
\prod_k[\frac{1}{k^2}]^{2(N^2-1)}[\frac{1}{k^2}]^{-(N^2-1)}\frac{1}{k^2}^{-2NN_f}]\approx[\exp[-iq(\frac{1}{32}+\frac{5}{64}\epsilon+...)].
\label{res72553}
\end{eqnarray}
where,
\begin{eqnarray}
q=2(N^2-1)-(N^2-1)-2N_fN.
\label{res52999097}
\end{eqnarray}
However the final result will be independent of the contribution in Eq. (\ref{res72553}) which will factorize out.

The partition function $Z_0$ depends through the renormalization constants $Z_3$, $Z_2$ and $Z_2'$ on the scale $\mu^2$.
We consider the corresponding renormalization constants as follows (note that these must inherit the condition of renprmalization in the background gauge field method:
\begin{eqnarray}
{\cal L}=-\frac{1}{4g_0^2Z_3\mu^{2\epsilon}}F^{a\mu\nu}F^a_{\mu\nu}
-Z_{2}^{\prime}\partial^{\mu}\bar{c}^a\partial_{\mu}c^a+Z_{2}^{\prime}f^{abc}A^c_{\mu}\partial^{\mu}\bar{c}^ac^b+\sum_f[iZ_2\bar{\Psi}_f\gamma^{\mu}\partial_{\mu}\Psi_f+Z_2A^a_{\mu}\bar{\Psi}_f\gamma^{\mu}t^a\Psi_f].
\label{lagrrenorm647565}
\end{eqnarray}
Here we took into account that according to our prescription in the dimensional regularization the fields $\bar{c}^a$, $c^a$ have mass dimension $\frac{d-2}{2}$, $\bar{\Psi}$ and $\Psi$ mass dimension $\frac{d-1}{2}$ but the field $A^a_{\mu}$ has mass dimension $1$. Then to reestablish the correct dimensionality one must introduce a factor $\frac{1}{\mu^2}$ in front of the pure gauge field part of  the Lagrangian in Eq. (\ref{lagrrenorm647565}) Moreover,
\begin{eqnarray}
&&A^b_{\rho}(x)=Z_3^{1/2}A^b_{\rho r}(x)
\nonumber\\
&&\Psi(x)=Z_2^{1/2}\Psi_r(x)
\nonumber\\
&&\bar{\Psi}=Z_2^{1/2}\bar{\Psi}_r
\nonumber\\
&&c^a(x)=(Z_2')^{1/2}c^a_r(x)
\nonumber\\
&&\bar{c}^a=(Z_2')^{1/2}\bar{c}^a_r,
\label{renromquant566}
\end{eqnarray}
and the subscript $r$ indicates the renormalized wave functions.

Then:
\begin{eqnarray}
&&\frac{ d Z_0}{d\ln\mu^2}=
=\prod_{f,i,a}\int d A^a_{\mu}d\bar{\Psi}_{fi}d\Psi_{fi}d\bar{c}_adc_a\Bigg[-i(\epsilon+\frac{\beta(g^2)}{g^2})[\int d^d x-\frac{1}{4\mu^{2\epsilon}}F^a_{\mu\nu}F^{a\mu\nu}]+
\nonumber\\
&&\int d^dx\frac{d A^b_{\rho}(x)}{d Z_3}\frac{d Z_3}{d\ln\mu^2}\frac{\delta iS_0}{\delta A^b_{\rho}(x)}+
\nonumber\\
&&\int d^dx\frac{d \Psi(x)}{d Z_3}\frac{d Z_3}{d\ln\mu^2}\frac{\delta iS_0}{\delta\Psi(x)}+
\int d^dx\frac{d \bar{\Psi}(x)}{d Z_2}\frac{d Z_2}{d\ln\mu^2}\frac{\delta iS_0}{\delta \bar{\Psi}(x)}+
\nonumber\\
&&\int d^dx\frac{d c^a(x)}{d Z_2'}\frac{d Z_2'}{d\ln\mu^2}\frac{\delta iS_0}{\delta c^a(x)}+
\int d^dx\frac{d \bar{c}^a(x)}{d Z_2}\frac{d Z_2}{d\ln\mu^2}\frac{\delta iS_0}{\delta\bar{c}^a(x)}\Bigg]\exp[iS_0],
\label{importform75649384}
\end{eqnarray}
where $S_0$ is the action in the absence of the background gauge field.

In order to compute the last three lines in Eq. (\ref{importform75649384}) we need the definitions:
\begin{eqnarray}
&&Z_g^2=Z_3^{-1}
\nonumber\\
&&g_0^2=Z_g^2g^2
\nonumber\\
&&\frac{d \ln Z_g^2}{d\ln\mu^2}=-\frac{\beta(g^2)}{g^2}
\nonumber\\
&&\frac{d\ln Z_2}{d\ln \mu^2}=\gamma_2
\nonumber\\
&&\frac{d \ln Z_2'}{d\ln\mu^2}=\gamma_2',
\label{defre54663}
\end{eqnarray}
where $\beta(g^2)$ is the beta function for the coupling constant, $\gamma_2$ is the anomalous dimension of the fermion wave function, $\gamma_2'$ is the anomalous dimension of the ghost wave function.
We will also use the Schwinger -Dyson equation which refers to the fields equation of motion in the quantum approach. This states generically (\cite{Schwinger}, \cite{Dyson}) that  for an arbitrary field $\phi$:
\begin{eqnarray}
\langle 0|T\int d^4 x \phi^a(x_1)\frac{\delta S_0}{\delta \phi_b(x)}|0\rangle=-i\langle 0|\int d^4 x \frac{\delta \phi^a(x_1)}{\delta \phi^b(x)}|0\rangle,
\label{dysonecswhinger43667882}
\end{eqnarray}
where differentiation is considered in the functional sense.  Then for example,
\begin{eqnarray}
&&\Bigg[\prod_{f,i,a}\int d A^a_{\mu}d\bar{\Psi}_{fi}d\Psi_{fi}d\bar{c}_adc_a\int d^d x A^a_{\rho}(x)\frac{\delta iS_0}{\delta A^a_{\rho}(x)}\exp[iS_0]\Bigg]=
\nonumber\\
&&\int d^d x \delta(0)4(N^2-1)Z_0=4(N^2-1)\int d^d x \int \frac{d^dp}{(2\pi)^4}Z_0.
\label{didf54664}
\end{eqnarray}
Similar relations apply to all fields in Eq. (\ref{importform75649384}). Next we need to determine the first term on the right hand side of  Eq. (\ref{importform75649384}). For that we write
\begin{eqnarray}
&&-\frac{1}{4g_0^2Z_3\mu^{2\epsilon}}F^a_{\mu\nu}F^{a\mu\nu}=S_0-\Bigg[-\partial^{\mu}\bar{c}^a\partial_{\mu}c^a+\bar{c}^af^{abc}A^c_{\mu}\partial^{\mu}\bar{c}^ac^b+
\sum_f[i\bar{\Psi}_f\gamma^{\mu}\partial_{\mu}\Psi_f+A^a_{\mu}\bar{\Psi}_f\gamma^{\mu}t^a\Psi_f]\Bigg].
\label{res66388899}
\end{eqnarray}
We start with:
\begin{eqnarray}
&&\prod_{f,i,a}\int d A^a_{\mu}d\bar{\Psi}_{fi}d\Psi_{fi}d\bar{c}_a dc^a\int d^d x A^a_{\rho}(x)\frac{\delta iS_0}{\delta A^a_{\rho}(x)}\exp[iS_0]=
\int d^d x\int \frac{d^dp}{(2\pi)^4}=
\nonumber\\
&&\prod_{f,i,a}\int d A^a_{\mu}d\bar{\Psi}_{fi}d\Psi_{fi}d\bar{c}_a dc^a \Bigg[\int d^dx \frac{\delta}{\delta A^a_{\rho}}[A^a_{\rho}(x)\exp[iS_0]]-
\int d^dx\frac{\delta A^a_{\rho}}{\delta A^a_{\rho}}iS_0\exp[iS_0]\Bigg]=
\nonumber\\
&&-i\int d^d x\int \frac{d^dp}{(2\pi)^4}\prod_{f,i,a}\int d A^a_{\mu}d\bar{\Psi}_{fi}d\Psi_{fi}d\bar{c}_a dc^a S_0\exp[iS_0].
\label{res5352442}
\end{eqnarray}
Here we applied the principle of integration by parts and $a$ and $\rho$ are considered fixed. Then Eq. (\ref{res5352442}) leads to:
\begin{eqnarray}
\prod_{f,i,a}\int d A^a_{\mu}d\bar{\Psi}_{fi}d\Psi_{fi}d\bar{c}_a dc^a S_0\exp[iS_0]=iZ_0,
\label{res6354663}
\end{eqnarray}
result which we shall use in what follows.

One can further use:
\begin{eqnarray}
&&\prod_{f,i,a}\int d A^a_{\mu}d\bar{\Psi}_{fi}d\Psi_{fi}d\bar{c}_a dc^a\sum_f[i\bar{\Psi}_f\gamma^{\mu}\partial_{\mu}\Psi_f+A^a_{\mu}\bar{\Psi}_f\gamma^{\mu}t^a\Psi_f]\exp[iS_0]=
\nonumber\\
&&\prod_{f,i,a}\int d A^a_{\mu}d\bar{\Psi}_{fi}d\Psi_{fi}d\bar{c}_a dc^a(-i)\frac{d}{d\ln Z_2}\exp[iS_0]=
\nonumber\\
&&-i2N_fN\int d^d x\int \frac{d^dp}{(2\pi)^4}Z_0,
\label{res55288990}
\end{eqnarray}
where we applied the procedure form Eqs. (\ref{importform75649384}) and (\ref{didf54664}). Similarly,
\begin{eqnarray}
&&\prod_{f,i,a}\int d A^a_{\mu}d\bar{\Psi}_{fi}d\Psi_{fi}d\bar{c}_a dc^a\Bigg[-\partial^{\mu}\bar{c}^a\partial_{\mu}c^a+f^{abc}A^c_{\mu}\partial^{\mu}\bar{c}^ac^b\Bigg]\exp[iS_0]=
\nonumber\\
&&-i\frac{1}{2}(N^2-1)\int d^d x\int \frac{d^dp}{(2\pi)^4}Z_0.
\label{ghost4665554}
\end{eqnarray}
Finally collecting the results in Eqs. (\ref{importform75649384}), (\ref{res66388899}), (\ref{res6354663}), (\ref{res55288990}) and (\ref{ghost4665554}) we obtain:
\begin{eqnarray}
&&\frac{dZ_0}{d\ln(\mu^2)}=
\nonumber\\
&&\Bigg[2(N^2-1)(\epsilon+\frac{\beta(g^2)}{g^2})-4N_fN\gamma_2-(N^2-1)\gamma_2'+(\epsilon+\frac{\beta(g^2}{g^2})[2N_fN+(N^2-1)/2]\Bigg]\int d^dx\int \frac{d^dp}{(2\pi)^4}Z_0.
\label{finalres663552442}
\end{eqnarray}
Here we shall use:
\begin{eqnarray}
\int d^d x\int \frac{d^dp}{(2\pi)^4}=-\frac{1}{4^{d/2}}(\frac{d}{2})^2(\gamma[-\frac{d}{2}])^2x^2=-[\frac{1}{16}+\frac{1}{8}\epsilon+...],
\label{calc65788477}
\end{eqnarray}
where we applied  Eq. (\ref{somecalc774665}) and the results in the Appendix.
Then one can write:
\begin{eqnarray}
&&\frac{dZ_0}{d\ln(\mu^2)}=Z_0\Bigg[(\epsilon+\frac{\beta(g^2)}{g^2})(u_0+u_1\epsilon)+(t_0+t_1\epsilon)\gamma_2+(s_01+s_1\epsilon)\gamma_2'\Bigg],
\label{reltobeused3443}
\end{eqnarray}
where $u_0$, $u_1$, $t_0$, $t_1$, $s_0$ and $s_1$ are dimensionless coefficients to be extracted from Eqs. (\ref{finalres663552442}) and (\ref{calc65788477}) whose exact values are irrelevant for what follows.

\section{The perturbative beta function}
We start with the final equation in section II, Eq (\ref{res664553}) which we shall rewrite here for completeness:
\begin{eqnarray}
\Bigg[\frac{1}{g^2}\mu^{2\epsilon}(\frac{r_1}{\epsilon}+r_0)+2b\frac{1}{g^4}\Bigg]Z_0=0.
\label{resqqq664553}
\end{eqnarray}
We apply the operator $\frac{d}{d\ln\mu^2}$ which yields:
\begin{eqnarray}
&&\Bigg[\epsilon\frac{1}{g^2}\mu^{2\epsilon}(\frac{r_1}{\epsilon}+r_0)-\frac{\beta(g^2)}{g^4}\mu^{2\epsilon}(\frac{r_1}{\epsilon}+r_0)-\frac{2b\beta(g^2)}{g^6}\Bigg]+
\nonumber\\
&&[\frac{1}{g^2}\mu^{2\epsilon}(\frac{r_1}{\epsilon}+r_0+...)+\frac{b}{g^4}]
\Bigg[(\epsilon+\frac{\beta(g^2)}{g^2})(u_0+u_1\epsilon+...)+(t_0+t_1\epsilon)\gamma_2+(s_0+s_1\epsilon)\gamma_2'\Bigg]=0,
\label{masterform64788}
\end{eqnarray}
where we used Eq. (\ref{reltobeused3443}).

Since $\epsilon$ is a small independent parameter we first equate the coefficient of $\frac{1}{\epsilon}$ to zero:
\begin{eqnarray}
-\frac{\beta(g^2)}{g^4}r_1+r_1u_0\frac{\beta(g^2)}{g^4}+\frac{1}{g^2}r_1(t_0\gamma_2+s_0\gamma_2')=0,
\label{res635552432}
\end{eqnarray}
from which we derive:
\begin{eqnarray}
t_0\gamma_2+s_0\gamma_2'=\frac{\beta(g^2)}{g^2}(1-u_0).
\label{anomdime73526635}
\end{eqnarray}
From Eq. (\ref{finalres663552442}) we determine:
\begin{eqnarray}
\frac{t_0}{t_1}=\frac{s_0}{s_1},
\label{exprofint65774}
\end{eqnarray}
which together with Eq. (\ref{anomdime73526635}) leads to:
\begin{eqnarray}
t_1\gamma_2+s_1\gamma_2'=\frac{t_1}{t_0}\frac{\beta(g^2)}{g^2}(1-u_0).
\label{res66253442}
\end{eqnarray}
Next we equate to zero the constant term in Eq. (\ref{masterform64788}) substitute the expressions in Eqs. (\ref{exprofint65774}) and (\ref{res66253442}) and solve for the beta function to obtain:
\begin{eqnarray}
\beta(g^2)=g^4r_1(1+u_0)\bigg[b-g^2[r_1u_1+r_1\frac{t_1}{t_0}(1-u_0)]\Bigg]^{-1}.
\label{re663554}
\end{eqnarray}
Then one can easily infer the exact expression for the beta function as:
\begin{eqnarray}
\beta(g^2)=-g^4\frac{\beta_0}{1-\frac{\beta_1}{\beta_0}g^2},
\label{betaufnc645538}
\end{eqnarray}
where $\beta_0$ and $\beta_1$ are the first two orders renormalization scheme independent coefficients.

\section{The non-perturbative beta function}

In this section we will show how the procedure employed in the previous section can be adjusted easily for the case
when the theta angle term is introduced in the left hand side and right hand side of the formula in Eq. (\ref{finalremaster4566}) and when $B^a_{\mu}$ can be assimilated to an arbitrary instanton solution.
For that $B^a_{\mu}$ must not only satisfy the equation of motion but also the condition:
\begin{eqnarray}
\int d^4 x\frac{1}{32\pi^2}\epsilon^{\mu\nu\rho\sigma}F^a_{\mu\nu}F^a_{\rho\sigma}=n
\label{cond66577}
\end{eqnarray}
where $F^a_{\mu\nu}$ is the background gauge field tensor and n is the winding number. We shall label the instanton solution corresponding to the quantum number $n$ by $(B^a_{\mu})_n$.  Each solution will depend on the instanton scale $\rho$. First we apply the quadratic and quadrilinear derivatives as before to the left hand side and right hand side of the Eq.
(\ref{finalremaster4566}). Because of the particularity of the functional derivatives the $\theta$ term will not contribute to the left hand side derivatives of $\Gamma[B_n]$. However the left hand side we will contain an extra term given by the $\exp[-\Gamma[B_n]]$. One should obtain for both sides more complicated relation that depend on the instanton solution. In the end we shall take the limit $B_n\rightarrow B_{n0}$ where $B_{n0}$ is the particular solution for which $F^{a\mu\nu}=sgn(n)\epsilon^{\mu\nu\rho\sigma}F^a_{\rho\sigma}$ and also take the limit $\rho=\infty$. To see how this works  we write the corresponding instanton solution for $n=1$ \cite{Novikov1}:
\begin{eqnarray}
&&A^a_{\mu}=2\eta_{a\mu\nu}(x-x_0)_{\nu}\frac{1}{(x-x_0)^2+\rho^2}
\nonumber\\
&&F^a_{\mu\nu}=-4\eta_{a\mu\nu}\frac{\rho^2}{((x-x_0)^2+\rho^2)^2}.
\label{res774665}
\end{eqnarray}
Here,
\begin{eqnarray}
\eta_{a\mu\nu}=
\Bigg[
\begin{array}{cc}
\epsilon_{a\mu\nu}&\mu,\nu=1,2,3\\
-\delta_{a\nu}&\mu=4\\
\delta_{a\mu}&\nu=4\\
0&\mu=\nu=4
\end{array}
\Bigg].
\label{eta437665788}
\end{eqnarray}
It turns out that this solution leads to a constant when the first order gauge invariant is integrated in the action in the limit $\rho\rightarrow \infty$ but tends to zero in the higher order gauge invariant or when the background gauge field does not appear in the right combination. Moreover the instanton solution can be set to zero in the right hand side of the master formula in Eq. (\ref{finalremaster4566}) because it never appears in the right combination such that to lead to condition  (\ref{cond66577}) and in the limit $\rho\rightarrow \infty$ all contributions go to zero. But the right hand side will depend on $\theta$ in $Z_0(\theta)$ where the background gauge field is set to zero. Then all our previous computations in section III apply as well here with the exception of the exponential of the instanton action that will appear on the left hand side of Eq. (\ref{finalremaster4566}) and the dependence on $\theta$ in $Z_0(\theta)$ on the right hand side. Relations between the bilinear and quadrilinear derivatives hold as well in the form:
\begin{eqnarray}
W_{2l}=xW_{1l},
\label{eqimport65774}
\end{eqnarray}
where this time:
\begin{eqnarray}
&&W_{2l}=Z_{inst}^{-1}W_{2r}Z_0(\theta)
\nonumber\\
&&W_{1l}=Z_{inst}^{-1}W_{2r}Z_0(\theta).
\label{sol5788}
\end{eqnarray}
Here,
\begin{eqnarray}
Z_{inst}^{-1}=\exp[\frac{8\pi^2|n|}{g^2}-i\theta n].
\label{inst463544}
\end{eqnarray}

Then Eq. (\ref{eqimport65774}) will become:
\begin{eqnarray}
\Bigg[\frac{1}{g^2}\mu^{2\epsilon}(\frac{r_1}{\epsilon}+r_0+...)+2b\frac{1}{g^4}\Bigg]Z_0(\theta)\exp[\frac{8\pi^2|n|}{g^2}-i n\theta]=0.
\label{resqqq66455334}
\end{eqnarray}

Before going further we need to estimate $\frac{d Z_0}{d\ln\mu^2}$ in the presence of the theta term.  With the renormalization conditions we imposed the only change occurs in Eq. (\ref{res66388899}) which will become:
\begin{eqnarray}
&&\frac{dZ_0(\theta)}{d\ln(\mu^2)}=
\nonumber\\
&&\Bigg[2(N^2-1)(\epsilon+\frac{\beta(g^2)}{g^2})-4N_fN\gamma_2-(N^2-1)\gamma_2'+(\epsilon+\frac{\beta(g^2}{g^2})[2N_fN+(N^2-1)/2]\Bigg]\int d^dx\int \frac{d^dp}{(2\pi)^4}Z_0.
\label{finalres66355244243}
\end{eqnarray}
Note that the right hand side according to the path integral formalism does not contain the theta term anymore.

We apply the operator $\frac{d}{d\ln\mu^2}$ to Eq. (\ref{resqqq66455334}):
\begin{eqnarray}
\frac{d}{d\ln\mu^2}\Bigg[ Z_{inst}^{-1}Z_0(\theta)[\frac{1}{g^2}\mu^{2\epsilon}(\frac{r_1}{\epsilon}+r_0+...)+2b\frac{1}{g^4}]\Bigg]=0
\label{res5534}
\end{eqnarray}
This further leads to:
\begin{eqnarray}
&&\Bigg[\epsilon\frac{1}{g^2}\mu^{2\epsilon}(\frac{r_1}{\epsilon}+r_0)-\frac{\beta(g^2)}{g^4}\mu^{2\epsilon}(\frac{r_1}{\epsilon}+r_0)-2b\frac{\beta(g^2)}{g^6}\Bigg]+
\nonumber\\
&&\Bigg[\frac{1}{g^2}\mu^{2\epsilon}(\frac{r_1}{\epsilon}+r_0+...)+2b\frac{1}{g^4}\Bigg][-\frac{8\pi^2|n|}{g^4}\beta(g^2)-in\beta(\theta)]+
\nonumber\\
&&\Bigg[\frac{1}{g^2}\mu^{2\epsilon}(\frac{r_1}{\epsilon}+r_0+...)+2b\frac{1}{g^4}\Bigg]\times
\nonumber\\
&&\Bigg[(\epsilon+\frac{\beta(g^2)}{g^2})(u_0+u_1\epsilon)+(t_0+t_1\epsilon)\gamma_2+
(s_0+s_1\epsilon)\gamma_2'\Bigg]\exp[-\frac{8\pi^2|n|}{g^2}+i n\theta]=0.
\label{mastereq8299877}
\end{eqnarray}
Here we used Eq. (\ref{finalres66355244243})and the fact that $Z_{inst}^{-1}$ compensate for the instanton dependent part in $Z_0(\theta)$.

First we equate to zero the imaginary coefficient proportional to $\frac{1}{\epsilon}$ (Note that the constant imaginary coefficient is undetermined up to factors of $2\pi$ so we do not have a definite constraint).
This leads to:
\begin{eqnarray}
\beta(\theta)=\frac{1}{n}\sin(n\theta)\Bigg[\frac{\beta(g^2)}{g^2}u_0+t_0\gamma_2+s_0\gamma_2')\Bigg]\exp[-\frac{8\pi^2|n|}{g^2}].
\label{res64553443}
\end{eqnarray}
Next we equate to zero the $\frac{1}{\epsilon}$ term in the real part of Eq. (\ref{mastereq8299877}) to obtain:
\begin{eqnarray}
t_0\gamma_2+s_0\gamma_2'=\frac{1}{\cos(n\theta)}\Bigg[\exp[\frac{8\pi^2|n|}{g^2}]-u_0\cos(n\theta)+\frac{8\pi^2|n|}{g^2}\exp[\frac{8\pi^2|n|}{g^2}]\Bigg]\frac{\beta(g^2)}{g^2}.
\label{res6635453}
\end{eqnarray}
Next we equate to zero the constant real part which yields:
\begin{eqnarray}
\beta(g^2)=
&&g^4r_1\Bigg[1+u_0\exp[-\frac{8\pi^2|n|}{g^2}]\cos(n\theta)\Bigg]\times
\nonumber\\
&&\Bigg[(b-\frac{r_1t_1}{t_0}8\pi^2|n|)-g^2[\frac{r_1t_1}{t_0}[1-u_0\exp[-\frac{8\pi^2|n|}{g^2}]\cos(n\theta)]+r_1u_1\exp[-\frac{8\pi^2|n|}{g^2}]\cos(n\theta)]\Bigg]^{-1}.
\label{betaufnctiom4663553}
\end{eqnarray}
Finally substituting Eq. (\ref{res6635453}) into Eq. (\ref{res64553443}) one obtains:
\begin{eqnarray}
\beta(\theta)=\frac{\tan(n\theta)}{n}[1+\frac{8\pi^2|n|}{g^2}]\frac{\beta(g^2)}{g^2}.
\label{res6635442}
\end{eqnarray}

\section{A solution to the strong CP problem}
We consider the infrared region of the beta function. One can integrate  Eq. (\ref{betaufnc645538}) between two points $1$ and $2$ to obtain:
\begin{eqnarray}
-\frac{1}{\beta_0}[\frac{1}{g_2^2}-\frac{1}{g_1^2}]-\frac{\beta_1}{\beta_0^2}[\ln(g_2^2)-\ln(g_1^2)]=\ln(\mu_1^2)-\ln(\mu_2^2).
\label{intgr6664554}
\end{eqnarray}
For $g_1$ and $\mu_1$ in Eq. (\ref{intgr6664554}) arbitrary finite coupling constant and scale we consider $\mu_2\approx 0$ which leads to $\infty$ on the right hand side. Then in order for the left hand side to be $\infty$, $g_2$ must be equal to zero which shows that in the infrared region the beta function behave such that after the coupling constant increases to a maximum value starts to decrease as the scale is decreasing attaining zero for a zero scale. This result is in excellent agreement with  other important results in the literature using different approaches \cite{Deur1}, \cite{Deur2}, \cite{Binosi}.

From  Eq. (\ref{res6635442}) one can determine:
\begin{eqnarray}
\frac{d \theta}{d \ln(\mu^2)}=\frac{\tan(\theta n)}{n}[\frac{1}{g^2}+\frac{8\pi^2n}{g^4}]\frac{dg^2}{d\ln(\mu^2)},
\label{expr663882967}
\end{eqnarray}
which integrated between two values $g_1$, $\mu_1$ and $g_2$, $\mu_2$ leads to:
\begin{eqnarray}
\ln(\sin(\theta_2n))-\ln(\sin(\theta_1n))=\ln(g_2^2)-\ln(g_1^2)-\frac{8\pi^2n}{g_2^2}+\frac{8\pi^2n}{g_1^2},
\label{finalexpr7108322}
\end{eqnarray}
We consider  $\theta_1$ and $g_1$ at  a scale where both are finite. Then in the infrared region where $g_2^2=0$, $\theta_2$ must be zero as well.

Consequently the effective theta angle  in the nonperturbative regime is approximately zero fact indicated also by the experiments \cite{Afach}. Thus our beta functions provide a clear solution to the strong CP problem. This results is maintained for the full QCD case because adding fermions to our theory does not alter in any way the major steps that we took in our derivation.

\section{Discussion and conclusions}
It is little known about the QCD beta function for the coupling constant  in the  non-perturbative region and even less about the behavior of the theta angle.  In this work we first computed the all order perturbative beta function for QCD with fermions in the fundamental representation in the background gauge field method to find out that there are similarities with the form proposed by Pica and Sannino in \cite{Sannino} but also differences. Specifically in the absence of fermions, for pure Yang Mills, beta function coincides exactly to that in \cite{Sannino} but in the presence of fermions has a simpler form and it is completely determined. Based on the global properties of the partition function in the background of an instanton field with the winding number $n$  we further determined the non-perturbative beta functions for both the coupling constant and the theta angle. For that we extrapolate known properties of the instanton solution with $n=1$ to arbitrary instantons or antiinstantons with winding numbers $n$.    We then solve the two beta functions in the non-perturbative regime to obtain that the effective theta angle in this region is approximately zero.

Our work relates well with similar approximate results obtained in the literature for the non-perturbative beta functions \cite{Callan}, \cite{Morozov} and has some common features to the $N=2$ supersymmetric beta functions calculated  by Seiberg in \cite{Seiberg3}.

\begin{appendix}
\section{}

Here we shall calculate explicitly the integrals in Eq. (\ref{fin564555}) in the dimensional regularization approach.
We denote:
\begin{eqnarray}
&&I_1=\int \frac{d^d p}{(2\pi)^d}
\nonumber\\
&&I_2=\int \frac{d^d p}{(2\pi)^d} p^2
\nonumber\\
&&I_3=\int \frac{d^d p}{(2\pi)^d}\frac{1}{p^2}.
\label{res5453534}
\end{eqnarray}
In the standard dimensional regularization approach these integrals are considered zero \cite{Collins}. However it is necessary to put these integrals in the context to see how one can reach such a result. This will show that in our method the contributions of these integrals are by far nontrivial. The integrals can be solved such that to have  an ultraviolet cut-off or an infrared regulator. We will opt for the latter approach and introduce  a small infrared regulator $m^2$. We use the master formula \cite{Collins} in the euclidean space,
\begin{eqnarray}
&&\int \frac{d^dp}{(2\pi)^d}\frac{(p^2)^{\alpha}}{(p^2+m^2)^{\beta}}=
\nonumber\\
&&\frac{\pi^d}{(2\pi)^d}m^{d+2\alpha-2\beta}
\frac{\Gamma[\alpha+\frac{d}{2}]\Gamma[\beta-\alpha-\frac{d}{2}]}{\Gamma[\frac{d}{2}]\Gamma[\beta]},
\label{mastintegr454}
\end{eqnarray}
and further write for the integrals in Eq. (\ref{res5453534}):
\begin{eqnarray}
&&I_1=\int \frac{d^d p}{(2\pi)^d }\frac{(p^2)^{\alpha}}{(p^2-m^2)^{\alpha}}
\nonumber\\
&&I_2=\int \frac{d^d p}{(2\pi)^d}\frac{(p^2)^{\alpha+1}}{(p^2-m^2)^{\alpha}},
\label{secer545343}
\end{eqnarray}
where $\alpha=1$ and further,
\begin{eqnarray}
I_3=\int \frac{d^d p}{(2\pi)^d}\frac{1}{p^2-m^2}.
\label{thirdint656}
\end{eqnarray}

We need to calculate:
\begin{eqnarray}
&&X_1=I_1^3
\nonumber\\
&&X_2=I_2^2
\nonumber\\
&&I_3=I_1I_2I_3.
\label{intry65747}
\end{eqnarray}
We shall consider $m^2$ small and finite and only in the end take the limit $m \rightarrow 0$. First it can be easily deduced using the properties of the $\Gamma$ functions that:
\begin{eqnarray}
&&X_1=-(I_3')^3m^{3d}
\nonumber\\
&&X_2=(I_3')^2m^{2d+4}
\nonumber\\
&&X_3=(I_3')^3m^{3d},
\label{prop8867}
\end{eqnarray}
where $I_3'$ corresponds to $I_3$ with the  factor dependent on $m$ extracted. Here $I_3^2$ has a divergence of order $\frac{1}{\epsilon^2}$. Hence we can divide Eq. (\ref{fires534442}) by $(I_3')^2$ since in order for a product between an infinite quantity and another quantity to be zero the second quantity must be necessarily equal to zero. Then the quantities of interest are:
\begin{eqnarray}
&&Y_1=\frac{X_1}{(I_3')^2}
\nonumber\\
&&Y_2=\frac{X_2}{(I_3')^2}
\nonumber\\
&&Y_3=\frac{X_3}{(I_3')^2}.
\label{qunat647882}
\end{eqnarray}
In the dimensional regularization scheme with $4=4-2\epsilon$ we get:
\begin{eqnarray}
&&Y_1=\frac{1}{16\pi^2\epsilon}+\frac{1}{16\pi^2}+...
\nonumber\\
&&Y_2=1
\nonumber\\
&&Y_3=-\frac{1}{16\pi^2\epsilon}-\frac{1}{16\pi^2}+...
\label{res7253900}
\end{eqnarray}
which yields:
\begin{eqnarray}
&&x_1=x_0=\frac{1}{16\pi^2}
\nonumber\\
&&z_1=z_0=-\frac{1}{16\pi^2}
\nonumber\\
&&y_0=1
\nonumber\\
&&y_1=0.
\label{coef4567892}
\end{eqnarray}

\end{appendix}

\end{document}